\documentclass[12pt]{iopart}

\usepackage{iopams}  
\usepackage{graphicx}
\usepackage{subfigure}
\usepackage{color}
\usepackage{algorithm} 
\usepackage{algpseudocode} 
\expandafter\let\csname equation*\endcsname=\relax
\expandafter\let\csname endequation*\endcsname=\relax
\usepackage{amsmath,amssymb,amsthm}
\usepackage{hyperref}
\usepackage{cite}

\begin{document}
\title{Forked Physics-Informed Neural Networks for Non-Markovian Open Quantum Dynamics and Control}

\author{Zhao-Wei Wang$^{1}$, Kai-Yu Yuan$^{2}$, Feng-Hua Ren$^{3}$, Zhao-Ming Wang$^{1,4,5,*}$ }

\address{
$^{1}$ College of Physics and Optoelectronic Engineering, Ocean University of China, Qingdao 266100, China\\
$^{2}$ China Mobile (Suzhou) Software Technology Co., Ltd, Suzhou 215163, China \\
$^{3}$ School of Information Management and School of Artificial Intelligence, Qingdao University of Technology, Qingdao 266520, China\\
$^{4}$ Engineering Research Center of Advanced Marine Physical Instruments and Equipment of Ministry of Education, Ocean University of China, Qingdao, China\\
$^{5}$ Qingdao Key Laboratory of Optics and Optoelectronics, Ocean University of China, Qingdao 266100, China\\
$^{*}$ Author to whom anycorrespondence should be addressed.}

\ead{renfenghua@qtech.edu.cn, wangzhaoming@ouc.edu.cn}

\vspace{10pt}
\begin{indented}
\item[]Jul 2026
\end{indented}

\begin{abstract}
Physics-informed neural networks (PINNs) provide a pathway to reunify the simulation and control of quantum systems, in which these two tasks are typically decoupled in traditional strategies. However, most work remains confined to Markovian environments. When applied to non-Markovian systems, standard PINN architectures fail to converge reliably due to multi-objective optimization conflicts arising from the coupled differential equations. To address this fundamental limitation, we extend our previously proposed forked PINN (FPINN) by incorporating a dedicated control branch. By decoupling the optimization objectives at the gradient level via selective gradient flow, our method turns a previously intractable multi-task optimization into a well-conditioned one, allowing simulation and control to be optimized jointly without compromise. Numerical simulations on a two-qubit Heisenberg XXX model confirm that our framework faithfully reproduces the features of non-Markovian dynamics, including decoherence and information backflow. Taking a state‑preparation task on the same model as an example, our FPINN achieves higher fidelity than gradient ascent pulse engineering, chopped random basis, and standard PINNs, with the advantage becoming more pronounced as the environment becomes more dissipative and more Markovian. The generated pulses are also noticeably smoother, which is advantageous for experimental implementation. Our framework thus provides a unified, end-to-end differentiable paradigm for simulation and control of open quantum systems, with potential implications for quantum computing, simulation, and control.
\end{abstract}
\maketitle

\section{Introduction}
The dynamics of open quantum systems exhibit far richer and more complex behavior than their closed counterparts, making them central to quantum information science~\cite{nielsen2010quantum,breuer2002theory}, with applications ranging from quantum computing~\cite{Koch_2016, PhysRevApplied.17.054018} and quantum communication~\cite{zammit2026memory,Gaidi2026,10042303} to quantum metrology~\cite{PhysRevLett.109.233601,PhysRevA.98.042116}. Accurate simulation of open quantum system dynamics provides a fundamental tool for predicting quantum dynamics~\cite{li2021pulse,PRXQuantum.3.020301}, while quantum control offers a direct means to achieve high-fidelity quantum state preparation~\cite{wzb3-dbg9,z2tj-cwzb}, quantum gate operations~\cite{PhysRevLett.114.200502,yang2025quantum}, and entanglement generation~\cite{senoo2026high,p3cy-8yjk}. Usually, these two tasks, simulation and control, are decoupled: simulation relies on numerical integrators~\cite{NIEGEMANN2012364, LIU2024109055} or deep-learning-based simulators~\cite{shn6-53wk,lin2021simulation,IEEE9861710}, while control depends on independent optimization algorithms such as gradient ascent pulse engineering (GRAPE)~\cite{10.1098/rsta.2011.0361}, chopped random basis (CRAB)~\cite{PhysRevA.84.022326, PhysRevLett.106.190501}, or certain deep-learning-based optimizers~\cite{zhang2019does,niu2019universal}. This decoupling not only increases computational complexity, but also severs the intrinsic connection between system evolution and control design.

Recently, physics-informed neural networks (PINNs)~\cite{raissi2019physics,karniadakis2021physics,PhysRevResearch.4.023155, de2025inverse,CaiShengze2021Physics,cuomo2022scientific,ullah2024physics} have opened a new avenue to address this challenge by encoding the governing physical laws directly into the loss function and leveraging automatic differentiation (AD)~\cite{paszke2017automatic} to compute derivatives. By incorporating control objectives into the loss additionally~\cite{PhysRevLett.132.010801,IEEE10847579,batra2026physics,lauten2025physics}, such networks can simultaneously perform simulation and control within a unified framework. Norambuena et al.~\cite{PhysRevLett.132.010801} employed PINNs to achieve joint dynamical simulation and optimal control function learning in Markovian open quantum systems. Zhang et al.~\cite{IEEE10847579} further combined PINNs with sample-based learning to realize robust control in uncertain quantum systems. These works demonstrate the great potential of PINNs in unifying simulation and control within a single framework. However, these studies are restricted to Markovian environments, yet real quantum systems often exhibit non-Markovian memory effects~\cite{PhysRevLett.129.140501,PhysRevResearch.6.033243} that require solving multiple coupled auxiliary operator equations~\cite{Link2023,PhysRevA.58.1699,PhysRevA.60.91,PhysRevLett.83.4909}. 

Standard PINN architectures struggle with such coupled systems due to multi-objective optimization conflicts~\cite{yang2025s,cao2025wbpinn,zhang2026physics}. Incorporating control tasks compounds the difficulty further: the loss function must accommodate additional control objectives~\cite{PhysRevLett.132.010801,IEEE10847579}, and the network must simultaneously output both dynamical quantities and control parameters. This is not merely a technical nuisance. In quantum computing, high-fidelity gate operations must be implemented despite realistic noise spectra that are often non-Markovian~\cite{PhysRevB.76.045218,odeh2025non}. In quantum simulation, memory effects need to be properly considered in accurately capturing complex environment-mediated interactions~\cite{PhysRevA.111.062206}. More generally, quantum control protocols that account for environmental memory can outperform their Markovian counterparts~\cite{PhysRevA.98.062118,e27121239}. However, the practical realization of these advantages has been hindered by the lack of a unified computational framework that simultaneously handles both tasks without compromise. 

To overcome these challenges, i.e., the failure of standard PINNs in non-Markovian coupled simulations and the exacerbation of optimization conflicts when control tasks are introduced, we extend our previously proposed forked PINN (FPINN) by incorporating a dedicated control branch~\cite{wang2026forked}. By assigning separate branches to the auxiliary operator, the density matrix, and the control fields, we physically decouple the optimization objectives that would otherwise interfere in a monolithic network. Crucially, the selective gradient flow ensures that each branch is primarily updated by its own loss, while cross-branch gradients remain small. This design transforms the originally intractable multi-task optimization into a well-conditioned one, where the control pulses are optimized under the dynamical constraints without destabilizing the simulation accuracy. The FPINN thus does more than apply PINNs to a new problem class; it introduces a principled architectural strategy for handling coupled multi-task learning in physical systems. The result is an end-to-end differentiable framework that simultaneously simulates non-Markovian dynamics and generates physically consistent control pulses.

We first validate FPINN on non-Markovian dynamics simulations under both time-independent and time-dependent Hamiltonians, where it accurately reproduces the Runge-Kutta 4th-order (RK4) reference dynamics. For control pulse design in a state preparation task, FPINN achieves fidelity comparable to GRAPE, CRAB, and standard PINNs in the weak-dissipation regime, while outperforming them significantly under strong dissipation and stronger Markovianity. The generated pulses are also smoother and thus more experimentally feasible. These results establish FPINN as a unified, differentiable, and efficient framework for the simulation and control of open quantum systems.

\section{Model}
The total Hamiltonian of the open quantum system reads $ H_{tot} = H_s + H_b + H_{int} $, where $ H_s $, $ H_b $, and $ H_{int} $ represent the system, bosonic bath, and interaction parts, respectively. Setting $ \hbar = 1 $, the bath Hamiltonian is $ H_b = \sum_k \omega_k b_k^\dagger b_k $, where $ \omega_k $ is the frequency of mode $ k $, and $ b_k $, $b_k^\dagger$ are the corresponding annihilation and creation operators, satisfying the commutation relation $ [b_k, b_k^\dagger] = 1 $. The interaction Hamiltonian is taken as $H_{int} = \sum_k(g_k^* L^\dagger b_k + g_k L b_k^\dagger)$, with $ L $ as the Lindblad operator and $ g_k $ the complex coupling strength to mode $ k $.
	
The quantum state diffusion (QSD) approach~\cite{PhysRevLett.83.4909} is a trajectory method tailored for non-Markovian open quantum system dynamics. One projects the global wave function onto the coherent state $\left|z\right\rangle$ of the bath, obtaining the system state $|\psi(t,z^*)\rangle=\langle z|\Psi(t,z^*)\rangle $, where $ z=\{z_k\} $ is a set of complex Gaussian stochastic variables. Functional differentiation with respect to the noise defines the ansatz operator $ O $ via $ \frac\delta{\delta z_t^*}|\psi(t,z^*)\rangle=O(t,s,z^*)|\psi(t,z^*)\rangle$, so that each noise history generates a continuous quantum trajectory. Averaging over all trajectories yields the exact non-Markovian master equation for the reduced density matrix
\begin{equation}
	\frac{\partial \rho}{\partial t}= -\mathrm{i}[H_s,\rho_s] + [L,\rho_s\overline{O}^\dagger] - [L^\dagger,\overline{O}\rho_s]
	\label{equ:1}
\end{equation}
with the auxiliary operator $\overline{O}=\int_{0}^{t}ds\:\alpha(t,s)O(t,s,z^*_t)$, whose kernel is the bath correlation function $\alpha(t,s) = \int_{}^{}d\omega J(\omega)e^{-i\omega(t-s)} $. For the Lorentz-Drude spectrum~\cite{ritschel2014analytic,WANG201078} the spectral density reads $J(\omega)=\frac{\Gamma}{\pi}\:\frac{\omega}{1+(\omega/\gamma)^{2}}$, where $\Gamma$ quantifies the system-bath coupling strength and $ \gamma $ is the characteristic frequency of the bath. The limit $ \gamma \to 0 $ corresponds to coloured noise and strong non-Markovianity, whereas $ \gamma \to \infty $ recovers the Markovian white-noise limit. The master equation Eq.~(\ref{equ:1}) is closed by
\begin{equation}
	\frac{\partial\overline{O}}{\partial t} =\frac{\Gamma\gamma}{2}L-\gamma\overline{O}+[-\mathrm{i}H_s-L^\dagger\overline{O},\overline{O}],
	\label{equ:2}
\end{equation}
which propagates the mean operator $ \overline{O}(t) $ self-consistently. Together, Eqs.~\eqref{equ:1} and \eqref{equ:2} completely determine the non-Markovian dynamics of the reduced density matrix. Further details are given in Appendix A.

\section{Forked Physics-Informed Neural Networks}
\subsection{Neural network architecture} 
\begin{figure}[htbp]
	\centering{\includegraphics[width=\columnwidth]{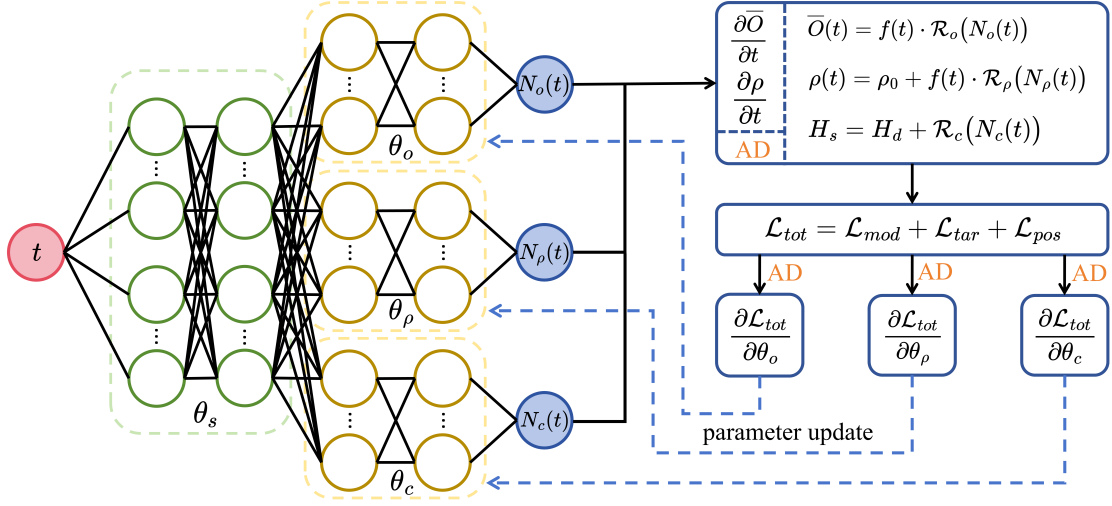}}
	\caption{FPINNs architecture for solving non-markovian open quantum dynamics and control. The input to FPINN is $t$. The green dashed box denotes the shared layers, which learn common features, while the yellow dashed box denotes the branch layers, which learn unique features. The outputs are feature vector $N_{o}(t)$, $N_{\rho}(t)$, and $N_{c}(t)$. Through reconstruction operations, $\overline{O}(t)$, $\rho(t)$, and $H_s(t)$ are obtained, and AD is used to compute $\partial\overline{O} /\partial t$ and $\partial\rho /\partial t$. After the gradients of the total loss with respect to the parameters of different branch layers are computed via AD, the updates are performed accordingly.} 
	\label{fig:1}
\end{figure}

As shown in Fig.~\ref{fig:1}, the FPINN adopts a shared trunk and multi‑branch parallel design to simultaneously predict three physical quantities, $\overline{O}(t)$, $\rho(t)$, and $c(t)$, from the time coordinate $t$. The network takes a scalar time $t=(t_{0},t_{1},\ldots,t_{f})$ as input, which first passes through two shared fully-connected layers $ \theta_s $: the first layer  maps the input to 512 dimensions, and the second maintains 512 dimensions. Each layer is followed by a SiLU activation and dropout regularization with a rate of 0.1, allowing the shared trunk to extract common temporal features while mitigating overfitting. The shared features are then fed into three independent branch networks $ \theta_o $, $ \theta_{\rho} $ and $ \theta_c $, respectively. Inside each branch, a linear layer reduces the 512‑dimensional features to 256 dimensions with a SiLU activation, and a second linear layer (256 → 256) with SiLU further enhances branch-specific nonlinear expressiveness. Finally, each branch applies an output linear layer to produce predictions of the target dimensions: the $\overline{O}$-branch outputs the auxiliary operator feature vector $N_o(t) \in \mathbb{R}^{32}$; the $\rho$-branch outputs the density matrix feature vector $N_\rho(t) \in \mathbb{R}^{15}$; and the $c$-branch outputs the control feature vector $N_c(t) \in \mathbb{R}^{4}$. All fully-connected layers in the network are initialized using Xavier uniform initialization (gain 0.05), and biases are sampled from a normal distribution $\mathcal{N}(0,0.01)$. With this design, the shared layers learn common temporal evolution patterns, while the branch networks capture the distinct complex mappings of different physical quantities, balancing parameter efficiency and multi-task learning flexibility.

\subsection{Loss function}
To enable FPINN to simulate open quantum system dynamics while simultaneously performing quantum control tasks, we construct a composite loss function consisting of three terms: the dynamics residual $ \mathcal{L}_{mod} $, the control target loss $ \mathcal{L}_{tar} $, and the density matrix positivity penalty $ \mathcal{L}_{pos} $: 
\begin{equation}
	\mathcal{L}_{tot} = \mathcal{L}_{mod} + \mathcal{L}_{tar} + \mathcal{L}_{pos}.
	\label{equ:3}
\end{equation}
The dynamics residual $ \mathcal{L}_{mod} $ consists of two sub-terms, $ \mathcal{L}_{mod}^{o} $ and $ \mathcal{L}_{mod}^{\rho} $, corresponding to the evolution of the auxiliary operator $ \overline{O}(t) $ and the density matrix $ \rho(t) $, respectively:
\begin{equation}
	\mathcal{L}_{mod}^{o} = \frac{1}{f}\sum_{i=0}^f \bigl\| \frac{\partial\overline{O}(t_i)}{\partial t} - \frac{\Gamma\gamma}{2}L + \gamma\overline{O}(t_i) -[-\mathrm{i}H_s-L^\dagger\overline{O}(t_i),\overline{O}(t_i)]\bigr\|^2,
	\label{equ:4}
\end{equation}
\begin{equation}
	\mathcal{L}_{mod}^{\rho} = \frac{1}{f}\sum_{i=0}^f \bigl\| \frac{\partial \rho(t_i)}{\partial t} + \mathrm{i}[H_s,\rho(t_i)]
	- \left[L,\rho(t_i)\overline{O}^\dagger(t_i)\right] + \left[L^\dagger,\overline{O}(t_i)\rho(t_i)\right]\bigr\|^2.
	\label{equ:5}
\end{equation}
Here $\left\|\cdot\right\|$ denotes the Euclidean norm. 

The control target loss drives the density matrix at the final time toward the target state
\begin{equation}
	\mathcal{L}_{tar} = \bigl\| \rho(t_f) - \rho_{tar} \bigr\|^2 .
	\label{equ:6}
\end{equation}
The positivity penalty $ \mathcal{L}_{pos} $ ensures the physical validity of the density matrix. Since $ \rho(t) $ must be positive semi-definite and Hermitian, we compute all negative eigenvalues of $ \rho(t) $ at each time step and take the sum of their squares
\begin{equation}
	\mathcal{L}_{pos} = \frac{1}{f}\sum_{n=1}^{f} \sum_{k=1}^{4} \bigl( \max(0, -\lambda_k(t_i) ) \bigr)^2 ,
	\label{equ:7}
\end{equation}
where $ \lambda_k(t_i) $ is the $ k $-th eigenvalue of $ \rho(t_i) $.

To strictly enforce the initial conditions, we employ hard constraints~\cite{PhysRevLett.132.010801} that map the raw network outputs to the physical quantities $\overline{O}(t)$ and $\rho(t)$. We define the temporal envelope function $f(t)=1-e^{-5t}$, which satisfies $f(t_0)=0$. The elements of the $4\times4$ complex matrix $\overline{O}(t)$ are obtained by arranging the 32 components of $N_o(t)$ in a fixed order as real and imaginary parts and then multiplying by $f(t)$, i.e., $\overline{O}(t)=f(t)\cdot\mathcal{R}_o\bigl(N_o(t)\bigr)$, where $\mathcal{R}_o$ denotes the reconstruction operation for $\overline{O}$, thus guaranteeing $\overline{O}(t_0)=0$ (the zero matrix). 

The reconstruction of the density matrix $\rho(t)$ exploits its Hermiticity and unit trace: the first three components of $N_\rho(t)$ directly give the diagonal elements $\rho_{11}$, $\rho_{22}$, $\rho_{33}$. The 4th to 14th components sequentially construct the upper-triangular elements (e.g., real and imaginary parts of $\rho_{12}$, $\rho_{13}$, etc.), each multiplied by $f(t)$. The lower-triangular elements are obtained via conjugate symmetry. Finally, trace normalization determines $\rho_{44}=1-\rho_{11}-\rho_{22}-\rho_{33}$. Thus $\rho(t)=\rho_0+f(t)\cdot\mathcal{R}_\rho\bigl(N_\rho(t)\bigr)$, where $\rho_0$ is the initial density matrix and $\mathcal{R}_\rho$ denotes the reconstruction mapping for $\rho$. At $t=0$, all terms multiplied by $f(t)$ vanish, exactly satisfying $\rho(t_0)=\rho_0$. This hard-constraint approach embeds the initial conditions directly into the network structure, eliminating the need for additional initial-value constraint terms in the loss function and avoiding the redundant parameters and initial-value errors that would arise from outputting each matrix element independently, as in some reference approaches.

For the control Hamiltonian, the output of the $c$-branch, $N_c(t)=[c_1(t),c_2(t),c_3(t),c_4(t)]$, directly provides the amplitudes of four independent control fields, and $\mathcal{R}_c(N_c(t))$ denotes the reconstruction mapping for $H_c$. Through this explicit construction, the raw network output $N_c(t)$ directly determines the time-dependent Hamiltonian $H_s(t)$, thereby seamlessly embedding the quantum control problem into the end-to-end differentiable framework.

\subsection{Training and Backpropagation}
In each training iteration, the time sampling point $t$ is fed forward through FPINN, yielding the raw outputs $N_o(t)$, $N_\rho(t)$, $N_c(t)$. The corresponding physical quantities are then obtained via the hard-constraint reconstructions: $\overline{O}(t)=f(t)\mathcal{R}_o(N_o(t))$, $\rho(t)=\rho_0+f(t)\mathcal{R}_\rho(N_\rho(t))$, and $H_s=H_d+\mathcal{R}_c(N_c(t))$. AD~\cite{paszke2017automatic} is used to compute the time derivatives $\partial\overline{O} /\partial t$, $\partial\rho/\partial t$, and the scalar total loss $\mathcal{L}_{tot}$ is evaluated according to the loss function defined earlier. The total loss comprises several contributions : the dynamics residuals from the $\overline{O}$-branch and the $\rho$-branch, while the terminal control loss $\mathcal{L}_{tar}$ and the positivity penalty $\mathcal{L}_{pos}$ depend solely on the $\rho$-branch. 

During backpropagation, each branch-specific loss term predominantly generates gradients that act on the parameters of that branch; cross-gradients to other branches exist due to physical coupling but are typically much smaller in magnitude. For instance, the dominant gradient for the $\overline{O}$-branch comes from $\mathcal{L}_{mod}^o$, while the $\rho$-branch receives its primary updates from $\mathcal{L}_{mod}^\rho$, $\mathcal{L}_{tar}$, and $\mathcal{L}_{pos}$. As these gradients propagate backward along their respective branches, they converge at the shared layers, where they are accumulated to update the shared parameters. See Appendix B for details.

Consequently, the shared layers integrate supervisory signals from all branches and are optimized to extract common temporal features; this is precisely the advantage of “hard parameter sharing” in multi-task learning. After backpropagation, the parameters are updated using the AdamW optimizer~\cite{loshchilov2017decoupled} (learning rate 0.005, weight decay $1\times10^{-5}$) with a cosine annealing learning rate schedule (minimum learning rate $1\times10^{-5}$). To prevent gradient explosion, gradient norms are clipped at 1.0 before the update. During training, the model parameters yielding the smallest validation loss are retained, and both the best and final models are saved. All computations are executed in parallel on a GPU.

\section{RESULTS AND DISCUSSIONS}
\subsection{Simulation of the Non-Markovian Open Quantum Dynamics}
We consider a two-qubit system with Hamiltonian $H_s$ decomposed into a fixed drift term $H_d$ and a designable control term $H_c$: 
\begin{equation}
	H_s = H_d + H_c.
	\label{equ:8}
\end{equation}
The drift term $H_d = J(\sigma_x\otimes\sigma_x + \sigma_y\otimes\sigma_y + \sigma_z\otimes\sigma_z)$ describes the isotropic Heisenberg (XXX) interaction between two qubits~\cite{nguyen2024programmable}, where $J = 0.5$ is the exchange coupling strength and $\sigma_{x,y,z}$ are Pauli matrices. This paradigmatic spin-1/2 model has been extensively used in quantum information science as a testbed for entanglement dynamics, quantum teleportation, and quantum simulation~\cite{Houça2022Entanglement, HOUCA2022169816, LOTSTEDT2024140975}. The control term $H_c$ can be flexibly chosen according to task requirements. We first consider a simple constant control scenario, applying a constant transverse field along the $x$-direction to each qubit: $H_c = \Omega(\sigma_x\otimes I + I\otimes\sigma_x)$, with $\Omega = 1$. The Lindblad operator $L = \sum_{i=1}^{2} \sigma_i^-$ represents collective dissipation, where $\sigma_i^-$ denotes the lowering operator~\cite{liang2020geometric}. The initial state is prepared as the product state $|\psi_0\rangle = |00\rangle$, i.e., $\rho_0 = |00\rangle\langle00|$. The system Hamiltonian $H_s$ is then time-independent, and the evolution is entirely determined by this static Hamiltonian. This part of the simulation does not require designing control pulses, so FPINN does not include a control branch.

\begin{figure}[htbp]
	\centering{\includegraphics[width=0.6\columnwidth]{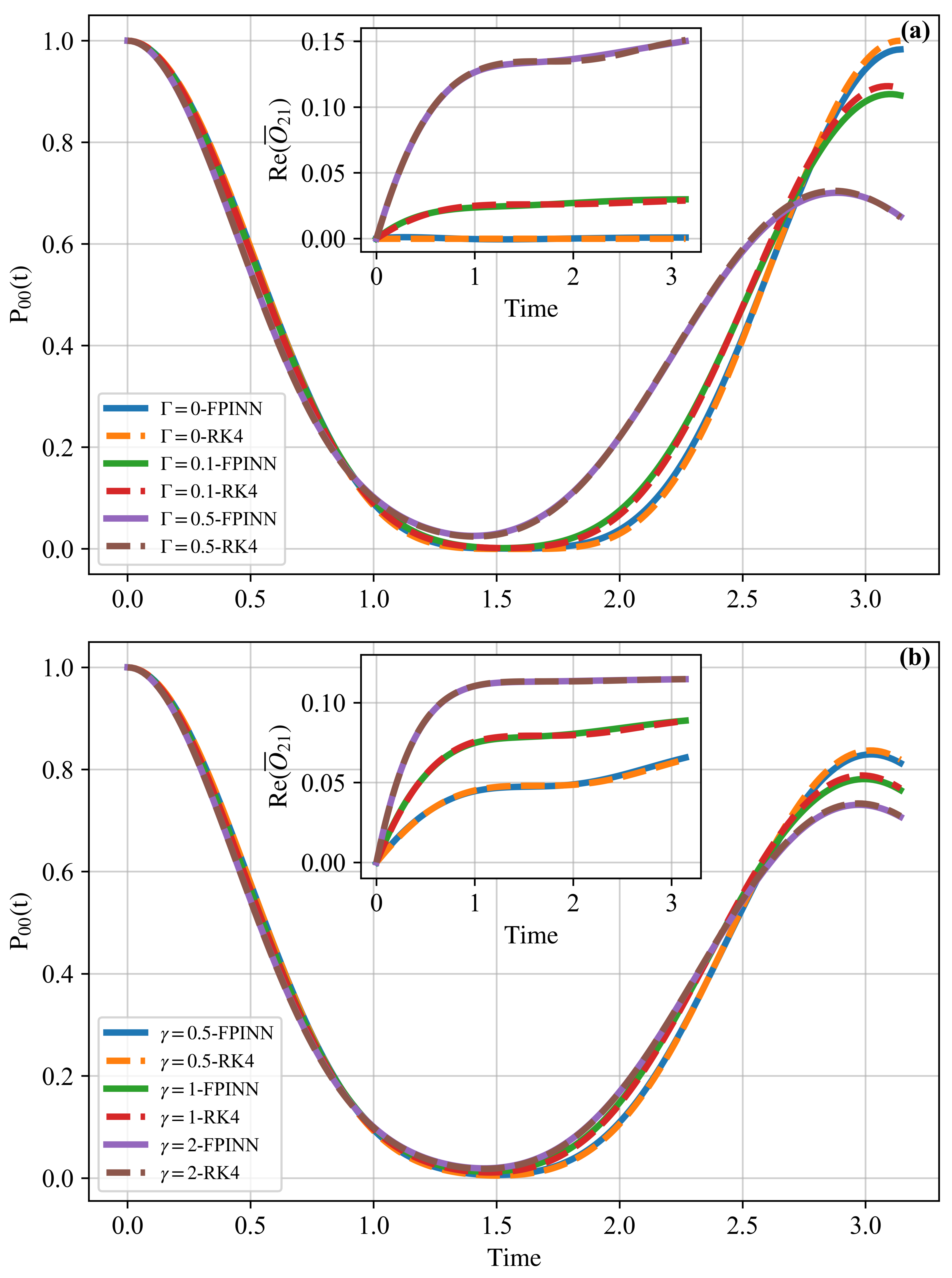}}
	\caption{(a) Evolution of $ P_{00}(t)$ simulated by RK4 and FPINN for different $ \Gamma $. (b) Evolution of $ P_{00}(t)$ simulated by RK4 and FPINN for different $ \gamma $.} 
	\label{fig:2}
\end{figure}

Under this time-independent Hamiltonian, we employ FPINN to simulate the open quantum dynamics governed by Eqs.~(\ref{equ:1}) and (\ref{equ:2}), including both coupling strength (characterized by $\Gamma$) and non-Markovian effects (characterized by the memory parameter $\gamma$). Figure~\ref{fig:2}(a) shows the ground-state population $P_{00}(t)=\langle00|\rho(t)|00\rangle $ simulated by FPINN for different values of the coupling strength $\Gamma$ with $\gamma=1$ fixed. When $\Gamma = 0$ (no environment), the system undergoes coherent oscillations and returns to its initial state at $t = \pi$ due to unitary evolution under $H_s$. As $\Gamma$ increases, the dissipative effect becomes stronger, and the ability to recover at $t = \pi$ gradually diminishes. FPINN predictions are in excellent agreement with the RK4 reference solution across all parameter settings.

Figure~\ref{fig:2}(b) examines the influence of parameter $\gamma$, which characterizes the memory time of the environment. A smaller $\gamma$ indicates stronger non-Markovianity (longer memory). The simulations fix $\Gamma = 0.3$ and vary $\gamma = 0.5, 1,$ and $2$. For smaller $\gamma$, the system exhibits a higher degree of recovery at $t = \pi$, a hallmark of non-Markovian dynamics where information flows back from the environment to the system. Conversely, larger $\gamma$ (approaching the Markovian limit) leads to weaker revivals. FPINN faithfully reproduces these features, matching the RK4 reference in all cases. The subplot in Fig.~\ref{fig:2} displays the real part of the off-diagonal density matrix element $\operatorname{Re}\{O_{12}\}$ as a function of time, which clearly captures the dephasing and decay trends.

\begin{figure}[htbp]
	\centering{\includegraphics[width=0.6\columnwidth]{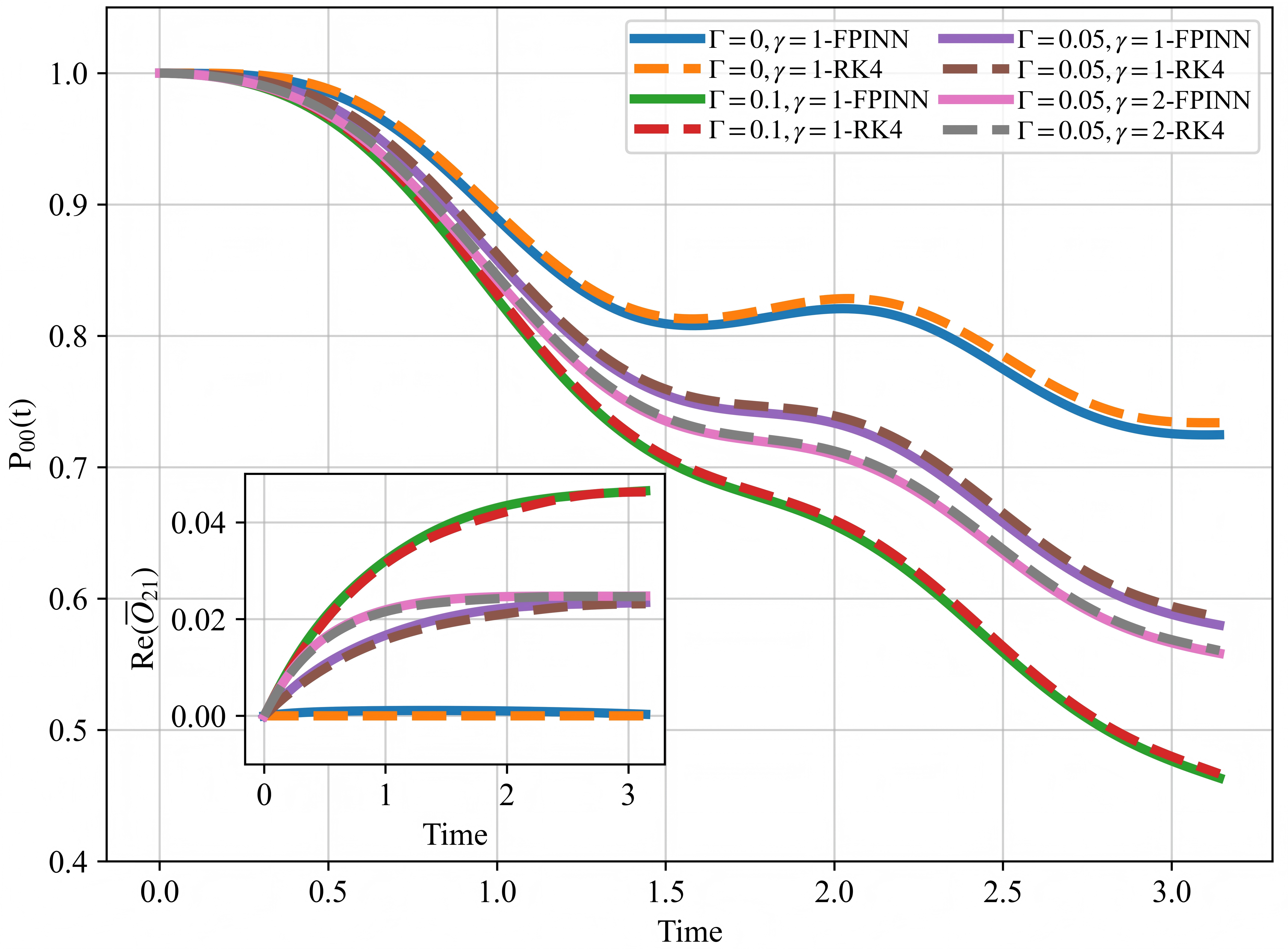}}
	\caption{Evolution of $ P_{00}(t)$ simulated by RK4 and FPINN for different $ \Gamma $ and $ \gamma $.} 
	\label{fig:3}
\end{figure}

We next consider a time-dependent Hamiltonian, where the control field is modulated sinusoidally in time. The control term now takes the form $H_c(t) = u_1(t)(\sigma_x\otimes I)$, with $u_1(t) = A\sin(\omega t)$, amplitude $A = 0.5$, and angular frequency $\omega = 2$. This time-dependent driving introduces a periodic modulation of the transverse field on the first qubit, while the system remains coupled to the environment. Compared with the time-independent case, the time-dependent Hamiltonian introduces additional complexity: the time-varying drive breaks the system's time-translation symmetry and introduces an external frequency scale $\omega$ that competes with the intrinsic coherent dynamics and the bath memory time $1/\gamma$, leading to richer but more challenging dynamics. Traditional numerical solvers require smaller time steps to resolve the oscillatory drive, which increases computational cost~\cite{PhysRevResearch.3.033266}. FPINN, however, bypasses this difficulty by treating $t$ as a continuous input to the network, so that the transition from a constant to a time-dependent Hamiltonian requires no modification of the network architecture. Figure~\ref{fig:3} presents the time evolution of $P_{00}(t)$, and the subplot displays the corresponding auxiliary operator $\overline{O}(t)$. The network faithfully reproduces the characteristic features of driven open quantum dynamics, including Rabi-type oscillations induced by the sinusoidal drive and the gradual decoherence due to environmental coupling. These observations demonstrate that FPINN is capable of high-fidelity simulation of non-Markovian open quantum dynamics even in the presence of a time-dependent Hamiltonian.

\subsection{FPINN for Quantum Control}
Beyond pure simulation, FPINN can also serve as a quantum control design tool by directly optimizing the control parameters in the time-dependent Hamiltonian. In this task, we consider the same two-qubit system with the drift Hamiltonian. Such a control architecture is realizable in superconducting qubit platforms, where microwave drives generate local $ \sigma_x $ and $ \sigma_y $ rotations on each qubit~\cite{hai2025scalable}, and the fixed Heisenberg coupling provides the entangling interaction~\cite{PhysRevB.78.064503,PhysRevB.79.024519}. The control Hamiltonian is constructed as a linear combination of four independent control fields: $H_c(t) = c_1(t)\,(\sigma_x\otimes I) + c_2(t)\,(I\otimes\sigma_x) + c_3(t)\,(\sigma_y\otimes I) + c_4(t)\,(I\otimes\sigma_y)$. The control coefficients $c_i(t)$ are the outputs of the FPINN $c$-branch, which is trained simultaneously with the $\overline{O}$ and $\rho$-branches. The objective is a quantum state preparation task: starting from the initial state $|\psi_0\rangle = |00\rangle$ (i.e. $\rho_0 = |\psi_0\rangle\langle\psi_0|$), the system should be driven to the target Bell state $|\phi^+\rangle = \frac{1}{\sqrt{2}}(|00\rangle + |11\rangle)$ (i.e. $\rho_{tar} = |\phi^+\rangle\langle\phi^+|$) at the final time $t_f$. The loss function includes the terminal fidelity term $\mathcal{L}_{tar} = \|\rho(t_f) - \rho_{tar}\|^2$ to enforce the control target.

\begin{figure}[htbp]
	\centering{\includegraphics[width=0.6\columnwidth]{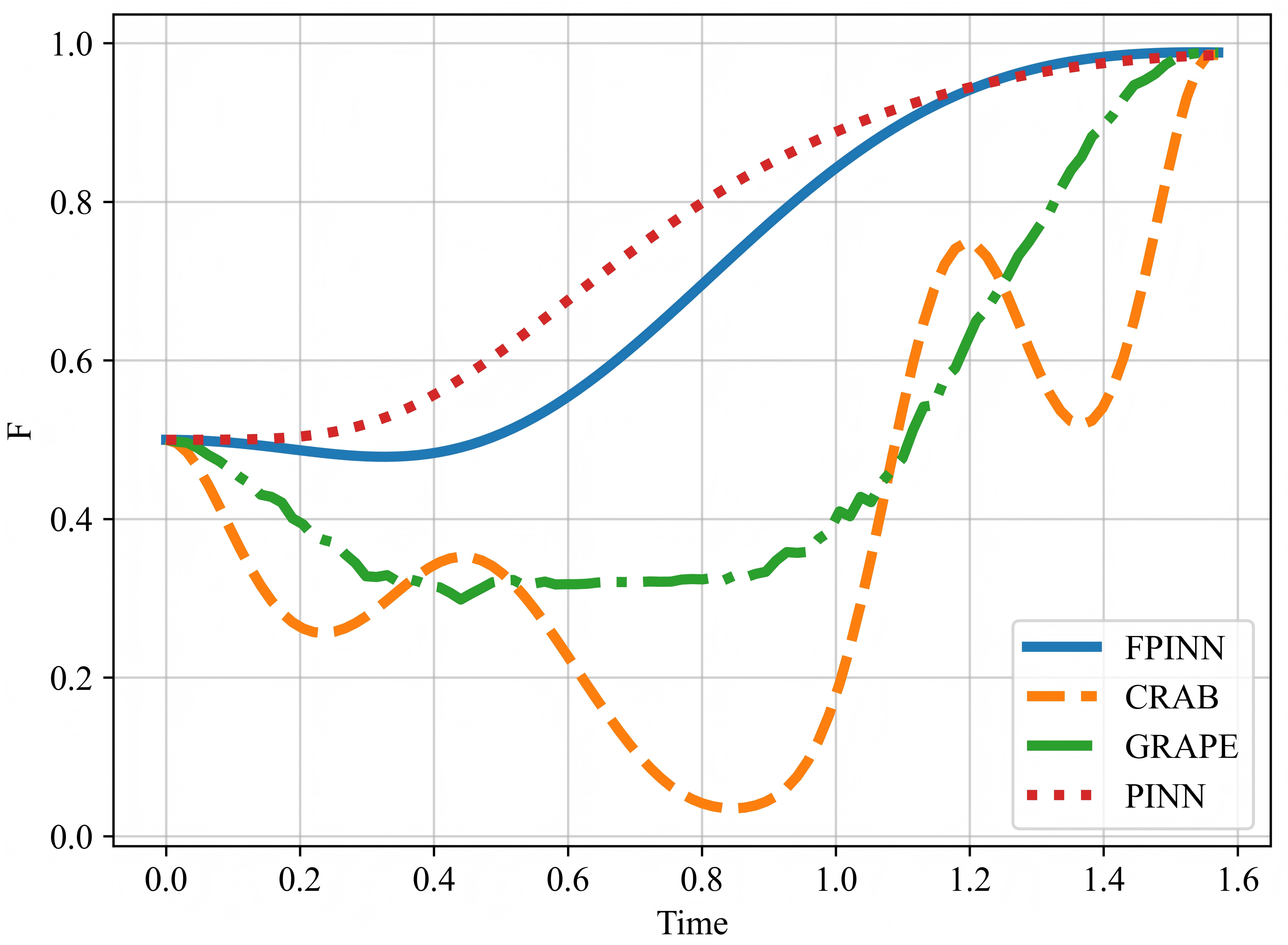}}
	\caption{Time evolution of the fidelity under the control schemes designed by the four algorithms.} 
	\label{fig:4}
\end{figure}

To demonstrate the effectiveness of FPINN in control pulse design, we compare it with three other methods: GRAPE~\cite{10.1098/rsta.2011.0361}, CRAB~\cite{PhysRevA.84.022326, PhysRevLett.106.190501}, and a standard PINN. 
The GRAPE and CRAB algorithms are implemented using the QuTiP package~\cite{Lambert_2026} with default parameter settings.
The standard PINN has the same number of hidden layers and neurons as FPINN, but without the branched architecture, it distinguishes different feature vectors only at the output layer~\cite{PhysRevLett.132.010801}. 
Figure~\ref{fig:4} shows the fidelity evolution (defined as $F = \operatorname{Tr}\sqrt{\sqrt{\rho_{tar}}\rho(t_i)\sqrt{\rho_{tar}}}$) for the control pulses designed by the four methods for the same environmental conditions ($\Gamma = 0.1$, $\gamma = 0.1$). The fidelity curves of FPINN and the standard PINN increase steadily, while those of GRAPE and CRAB exhibit larger fluctuations, indicating that the PINN-based methods yield more stable controls. The maximum fidelities achieved by FPINN, GRAPE, CRAB, and the standard PINN are 0.9886, 0.9881, 0.9855, and 0.9850, respectively, with FPINN attaining the highest value.

\begin{figure}[htbp]
	\centering{\includegraphics[width=0.6\columnwidth]{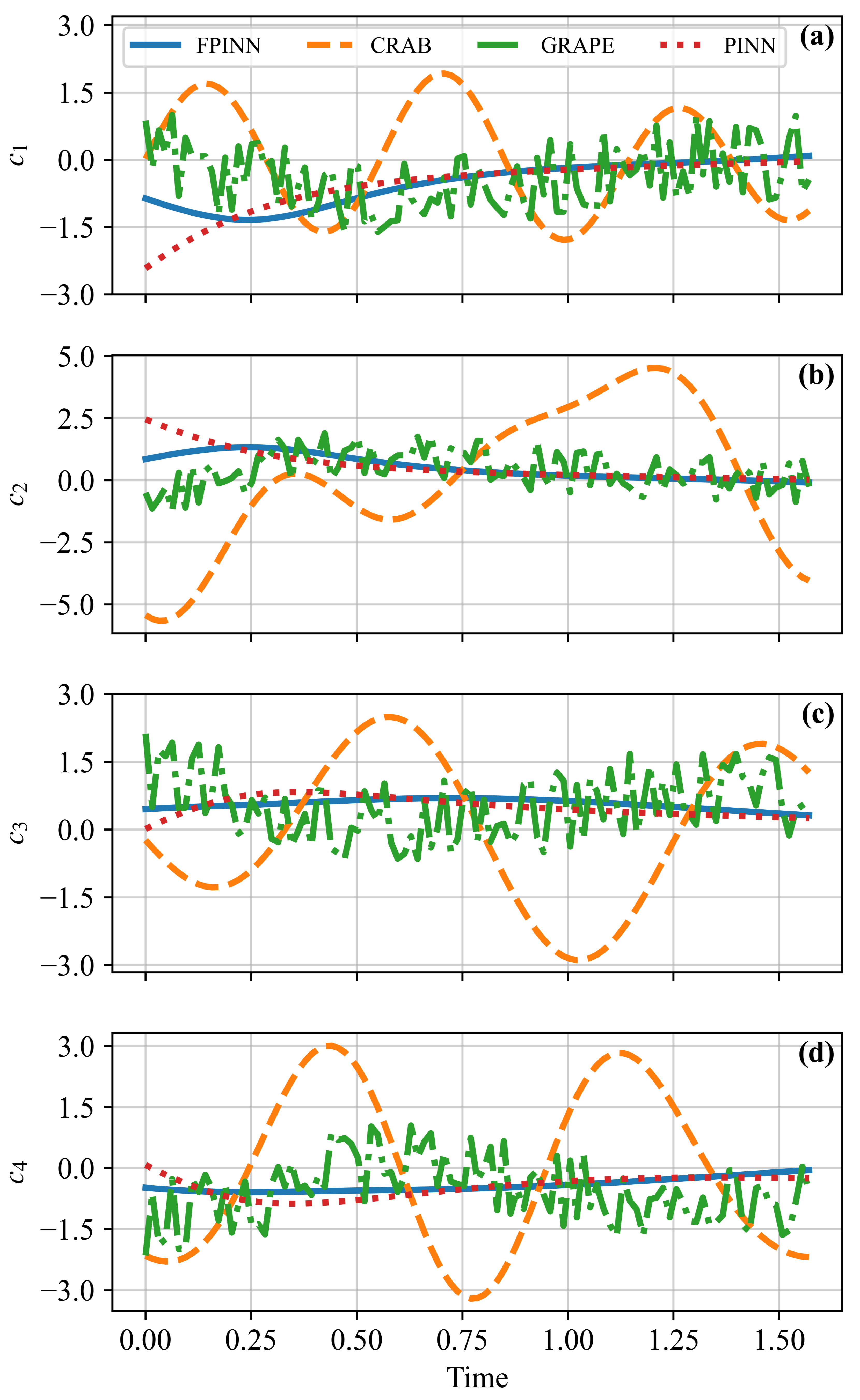}}
	\caption{Control amplitudes for the control schemes designed by the four algorithms.} 
	\label{fig:5}
\end{figure}

Figure~\ref{fig:5} shows the time evolution of the four control fields $c_1(t), c_2(t), c_3(t), c_4(t)$ obtained by the four methods. The pulses designed by FPINN and the standard PINN are noticeably smoother than those produced by GRAPE and CRAB, exhibiting smaller variations between adjacent time points and avoiding abrupt jumps or high-frequency oscillations. 
Table~\ref{tab:1} lists the peak-to-peak variation (maximum minus minimum) of each control field for the four methods. FPINN consistently produces the smallest variations across all four pulse components, with the largest range being 1.3552 for $c_2(t)$, which is smaller than the ranges observed for GRAPE and CRAB. The standard PINN also yields relatively smooth pulses, with slightly larger ranges than those of FPINN. These quantitative results further indicate that the control pulses generated by FPINN are the smoothest.
This smoothness property is particularly desirable for experimental implementation, as physical platforms such as superconducting qubits and trapped ions are subject to bandwidth constraints and finite rise times~\cite{song2026structured}. The FPINN framework naturally produces smooth pulses, which is a direct consequence of its continuous, differentiable network architecture. Through training via AD, the architecture implicitly regularizes the control parameters along the time dimension. As a result, FPINN not only achieves high-fidelity state preparation in complex non-Markovian environments, but also generates control pulses that are well-suited for direct experimental deployment.

\begin{table}[htbp]
	\centering
	\caption{Peak-to-peak variation of control fields for different methods.}
	\begin{tabular*}{\columnwidth}{@{\extracolsep{\fill}}c cccc}
		\hline
		Method & $c_1$ & $c_2$ & $c_3$ & $c_4$ \\
		\hline
		FPINN & 1.3513 & 1.3552 & 0.4647 & 0.3873 \\
		CRAB  & 3.7131 &10.1847 & 5.3839 & 6.2166 \\
		GRAPE & 2.6153 & 3.1039 & 2.7747 & 3.1947 \\
		PINN  & 2.3747 & 2.4137 & 0.8632 & 0.9046 \\
		\hline
	\end{tabular*}
	\label{tab:1}
\end{table}

\begin{figure}[htbp]
	\centering{\includegraphics[width=0.6\columnwidth]{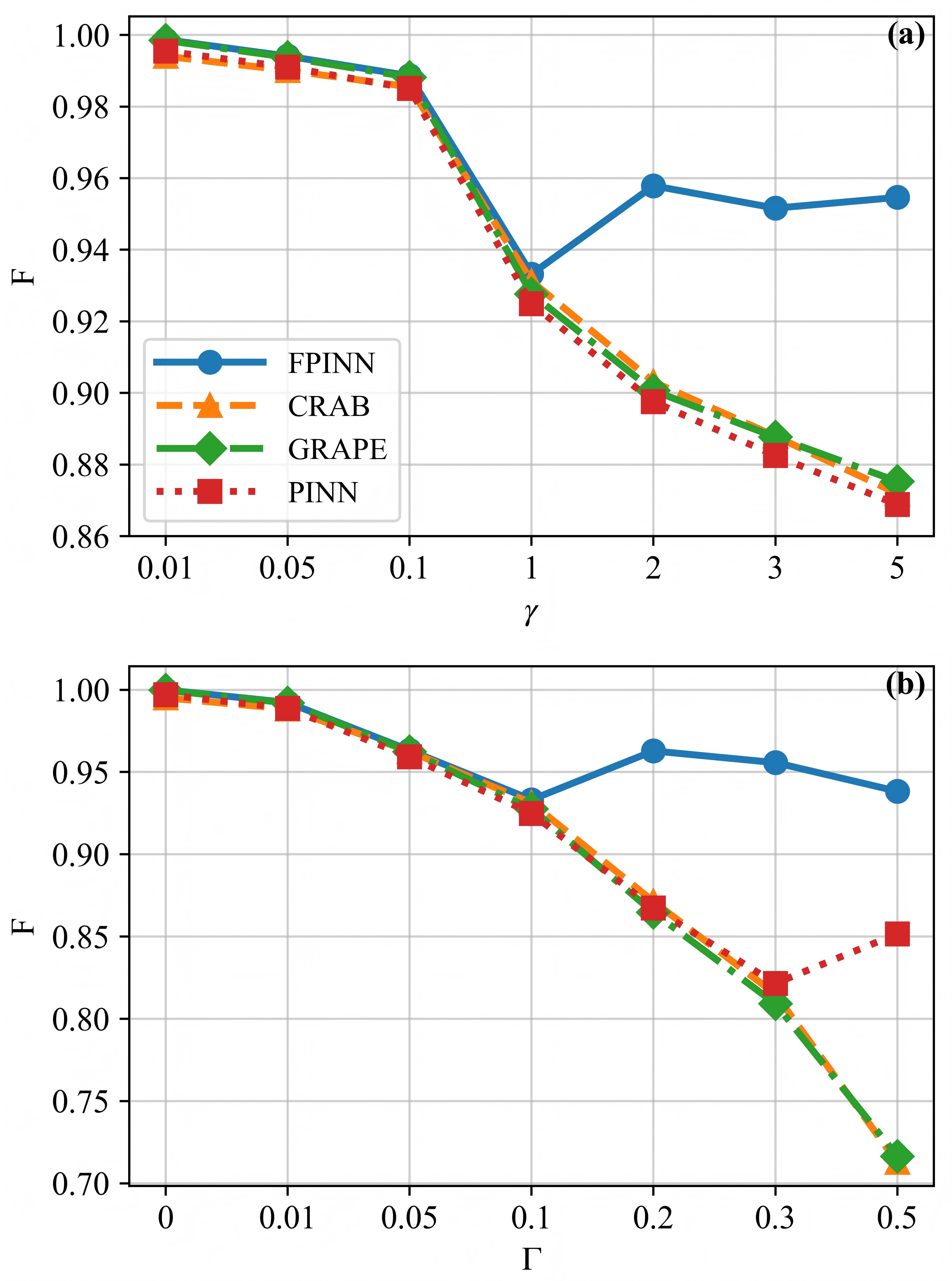}}
	\caption{Fidelities of the four methods for the same state-preparation problem under different environmental conditions. (a) $\gamma (\Gamma=0.1)$, (b) $\Gamma (\gamma=1)$.} 
	\label{fig:6}
\end{figure}

To further illustrate the practical advantage of FPINN in control design, we compare all four methods on the same state-preparation problem under various environmental conditions. Figure~\ref{fig:6}(a) shows the achieved fidelity as a function of the non-Markovian parameter $\gamma$ at a fixed moderate coupling strength ($\Gamma=0.1$). In the strong non-Markovian regime (small $\gamma$), all four algorithms achieve relatively high fidelities. This is because non-Markovian environments, characterized by long memory times, allow information to flow back from the bath to the system. This backflow, which manifests as revivals of coherence, can be harnessed by properly designed control pulses to counteract dissipation~\cite{PhysRevA.98.062118}. However, exploiting this memory effect requires the control algorithm to account for the entire history of the system-bath interaction. FPINN naturally incorporates this through its auxiliary operator $\overline{O}$, which encodes the environmental memory and is learned simultaneously with the control pulses. The standard PINN, in contrast, lacks a dedicated structure for handling the coupled dynamics, making it less effective at utilizing the memory information for control optimization. As $\gamma$ increases, the memory time shortens and the environment approaches the Markovian limit; the backflow diminishes, and the control performance of all methods degrades to some extent, with FPINN still outperforming the others due to its ability to adapt to the residual memory.

Figure~\ref{fig:6}(b) presents the fidelity versus the coupling strength $\Gamma$ (with $\gamma=1$ for a near-Markovian environment). At small coupling strengths, all four methods achieve high fidelities, which confirms the reliability of the FPINN approach. As $\Gamma$ grows, the coupling to the environment strengthens, making state preparation more challenging; all methods show a decreasing trend in fidelity. However, FPINN consistently maintains an advantage over GRAPE and CRAB, especially in the strong-coupling regime (e.g., $\Gamma \ge 0.1$). This is because FPINN learns, in a fully differentiable manner, how to actively shape the control pulses to counteract environmental disturbances. The network adjusts the $c(t)$ pulses to exploit information backflow in non-Markovian environments and to suppress decoherence in Markovian ones. The standard PINN exhibits a similar trend, but its performance degrades faster than FPINN as $\Gamma$ increases, highlighting the advantage of the branched architecture. CRAB and GRAPE, in contrast, lack the flexibility to adapt control pulses under strong dissipation, as they are not designed within a differentiable simulation-control unified framework. These results demonstrate that FPINN is not only a powerful simulator for open quantum dynamics but also an effective controller for high-fidelity quantum state preparation under both Markovian and non-Markovian environments.

\section{Conclusions}
We have proposed an FPINN for simultaneously simulating and controlling open quantum systems in non-Markovian environments. The network adopts a shared trunk and three branches to predict the evolution operator $\overline{O}(t)$, the density matrix $\rho(t)$, and the control pulses $c(t)$ from the time coordinate $t$. Its forked architecture and selective gradient flow allow each branch to be primarily driven by its own loss, while the shared layers aggregate multi-task gradients to extract common temporal features. FPINN accurately simulates non-Markovian dynamics under both time-independent and time-dependent Hamiltonians, matching RK4 reference results across a range of dissipation rates $\Gamma$ and memory parameters $\gamma$. In a quantum state preparation task, FPINN outperforms GRAPE, CRAB, and standard PINN under strong dissipation and Markovian conditions, as its jointly learned auxiliary operator $\overline{O}(t)$ encodes and exploits environmental memory. Moreover, the pulses produced by FPINN are significantly smoother, making them more amenable to experimental implementation. 

Beyond the two-qubit demonstration, the FPINN framework is readily generalizable to larger systems and more complex environments, as its computational cost scales with the dimension of the network rather than the dimension of the Hilbert space. This feature makes it particularly relevant for noisy intermediate-scale quantum devices, where reliable control under realistic non-Markovian noise remains a critical challenge. By providing a unified differentiable platform that integrates simulation and control, our approach offers a promising route toward optimal design of robust quantum gates and entanglement generation protocols in the presence of environmental memory. More broadly, the architectural principle of using selective gradient flow to decouple coupled optimization objectives is not limited to quantum systems; it can be applied to any multi-task learning problem governed by coupled differential equations, for example, in chemical dynamics or fluid turbulence \cite{cai2021physics}, where similar coupled equation structures arise.

\section{ACKNOWLEDGMENTS}
This work is supported by the Key R\&D Program of Shandong Province (Grant No. 2023CXGC010901), the Natural Science Foundation of Shandong Province (Grant No. ZR2024MA046), and the Fundamental Research Funds for the Central Universities (Grant No. 3002000-842661013) through the Graduate Independent Research Project of Ocean University of China.

\appendix
\section{Derivation of the non-Markovian master equation}
\label{app: cal_srd}
Within the QSD framework the system wave function is obtained by projecting the total state onto the Bargmann coherent state of the bath~\cite{PhysRevLett.83.4909}:
\begin{equation}
	\frac{\partial}{\partial t}|\psi(t,z_t^*)\rangle=[-iH_s+Lz_t^*-L^\dagger\overline{O}(t,z_t^*)]|\psi(t,z_t^*)\rangle,
	\label{equ:a1}
\end{equation}
where $z_t^*$ is complex Gaussian noise,$ \overline{z_{t}^*}=0 $, $\overline{z_{t}^*z_{s}}=\delta(t-s)$. The operator $ \overline{O} $ is defined as
\begin{equation}
	\overline{O}(t,z_t^*)=\int_{0}^{t}ds\:\alpha(t,s)O(t,s,z_t^*),
	\label{equ:a2}
\end{equation}
with the functional-derivative ansatz
\begin{equation}
	O(t,s,z^*)|\psi(t,z^*_t)\rangle=\frac\delta{\delta z_t^*}|\psi(t,z^*_t)\rangle.
	\label{equ:a3}
\end{equation}
The bath correlation function
\begin{equation}
	\alpha(t,s) = \int_{}^{}d\omega J(\omega)e^{-i\omega(t-s)},
	\label{equ:a4}
\end{equation}
where $J(\omega)=\frac{\Gamma}{\pi}\frac{\omega}{1+(\omega/\gamma)^{2}}$ is the bath spectral density.
Consistency requires
\begin{equation}
	\frac{\partial \overline{O}}{\partial t}=[-iH_s+Lz_t^*-L^\dagger\overline{O},O] -L^\dagger\frac{\delta\overline{O}}{\delta z_s^*}.
	\label{equ:a5}
\end{equation}
The reduced density matrix is the ensemble average 
\begin{equation}
	\rho_s=M[P_t]=M[|\psi(t,z_t^*)\rangle\langle\psi(t,z_t^*)|],
	\label{equ:a6}
\end{equation}
with $ M[\cdot]=\prod_k\frac1\pi\int d^2ze^{-|z|^2}(\cdot)$.
Taking the time derivative gives the exact non-Markovian master equation
\begin{equation}
	\begin{split}
		\frac{\partial\rho_s}{\partial t}=&-i[H_s,\rho_s]+[L,M[P_t\overline{O}^\dagger(t,z_t^*)]]\\
		&+[M[\overline{O}(t,z_t^*)P_t],L^\dagger]
	\end{split}
	\label{equ:a7}
\end{equation}
The $ O $ operator containing stochastic noise $z_t^*$ is usually obtained approximately by perturbative techniques
\begin{equation}
	\begin{split}
		O(t,s,z_t^*)
		=&O_0(t,s)+\int_0^tO_1(t,s,s_1)z_{s_1}ds_1+ \cdots\\ 
		&+\int_0^t\cdots\int_0^tO_n(t,s,\cdots,s_n)z_{s_1}\cdots z_{s_n} ds_1\cdots ds_n + \cdots
	\end{split}
	\label{equ:a8}
\end{equation}
and for weak coupling we retain only the zeroth-order term, $O(t,s,z_{t}^{*})\approx O_{0}(t,s)\equiv O(t,s)$. Within this approximation, $ M[P_t \overline{O}^\dagger(t, z_t^*)] = \rho_s \overline{O}(t) $, and Eq.~(\ref{equ:1}) reduces to the closed non-Markovian master equation quoted in the main text.

\section{Backpropagation and gradient flow in FPINN}
The backpropagation process follows the chain rule, with gradients flowing through the computational graph. The shared layer parameters $\theta_s$ comprise the weights and biases of the two shared fully connected layers (mapping the input time $t$ to a common feature representation). The branch parameters $\theta_o$, $\theta_\rho$, and $\theta_c$ include the weights and biases of the two hidden layers and the output layer within each respective branch. The branches produce the raw vectors $N_o(t)$, $N_\rho(t)$, and $N_c(t)$. Using the time envelope $f(t)=1-e^{-5t}$, the physical quantities are reconstructed as $ \overline{O}(t)=f(t)\mathcal{R}_o(N_o(t)) $, $ \rho(t)=\rho_0+f(t)\mathcal{R}_\rho(N_\rho(t)) $ and $ H_{s}=H_{d}+\mathcal{R}_{c}(N_{c}(t)) $, where $\mathcal{R}_o$, $\mathcal{R}_\rho$, and $\mathcal{R}_c$ denote the reconstruction mappings for the auxiliary operator, the density matrix, and the control Hamiltonian, respectively.

The total loss function is
\begin{equation}
	\mathcal{L}_{tot} = \mathcal{L}_{mod}^o(\theta_s,\theta_o,\theta_c) + \mathcal{L}_{mod}^\rho(\theta_s,\theta_\rho,\theta_o,\theta_c)  + \mathcal{L}_{tar}(\theta_s,\theta_\rho) + \mathcal{L}_{pos}(\theta_s,\theta_\rho),
	\label{equ:b1}
\end{equation}
where the dependencies on the network parameters are explicitly indicated. Here $\mathcal{L}_{mod}^o$ enforces the differential equation for $\overline{O}(t)$ and thus depends on $\overline{O}$ and $H_s$, i.e. on $\theta_o$, $\theta_c$, and $\theta_s$. $\mathcal{L}_{mod}^\rho$ enforces the Lindblad master equation for $\rho(t)$ and depends on $\rho$, $\overline{O}$, and $H_s$, hence on $\theta_\rho$, $\theta_o$, $\theta_c$, and $\theta_s$. $\mathcal{L}_{tar}$, and $\mathcal{L}_{pos}$ depend only on $\rho(t)$, i.e. only on $\theta_\rho$ and $\theta_s$.

During backpropagation, the gradient of the total loss with respect to the parameters of each branch is obtained by summing the gradients of all loss terms that depend on that branch. Specifically,
\begin{equation}
	\begin{aligned}
		\frac{\partial \mathcal{L}_{tot}}{\partial \theta_o} &= \frac{\partial \mathcal{L}_{mod}^o}{\partial \theta_o} + \frac{\partial \mathcal{L}_{mod}^\rho}{\partial \theta_o}, \\[4pt]
		\frac{\partial \mathcal{L}_{tot}}{\partial \theta_\rho} &= \frac{\partial \mathcal{L}_{mod}^\rho}{\partial \theta_\rho} + \frac{\partial \mathcal{L}_{tar}}{\partial \theta_\rho} + \frac{\partial \mathcal{L}_{pos}}{\partial \theta_\rho}, \\[4pt]
		\frac{\partial \mathcal{L}_{tot}}{\partial \theta_c} &= \frac{\partial \mathcal{L}_{mod}^o}{\partial \theta_c} + \frac{\partial \mathcal{L}_{mod}^\rho}{\partial \theta_c}.
	\end{aligned}
	\label{equ:b2}
\end{equation}
The shared layer parameters $\theta_s$ receive gradients from all loss components
\begin{equation}
	\frac{\partial \mathcal{L}_{tot}}{\partial \theta_s} = \frac{\partial \mathcal{L}_{mod}^o}{\partial \theta_s} + \frac{\partial \mathcal{L}_{mod}^\rho}{\partial \theta_s} + \frac{\partial \mathcal{L}_{tar}}{\partial \theta_s} + \frac{\partial \mathcal{L}_{pos}}{\partial \theta_s}.
	\label{equ:b3}
\end{equation}
These expressions reflect the actual coupling between the branches: the $\overline{O}$-branch receives gradients from both $\mathcal{L}_{mod}^o$ and $\mathcal{L}_{mod}^\rho$ because the density matrix equation explicitly uses $\overline{O}(t)$. The $\rho$ branch receives gradients only from $\mathcal{L}_{mod}^\rho$, $\mathcal{L}_{tar}$, and $\mathcal{L}_{pos}$; it does not receive gradients from $\mathcal{L}_{mod}^o$ because $\mathcal{L}_{mod}^o$ does not involve $\rho(t)$. The $c$-branch $\theta_c$ is shared by both dynamical equations and therefore accumulates gradients from $\mathcal{L}_{mod}^o$ and $\mathcal{L}_{mod}^\rho$. The shared layers $\theta_s$ aggregate all gradients, enabling them to learn features that benefit all three tasks.

For the $\overline{O}$-branch parameters $\theta_o$, the full chain-rule expansion of the total gradient is
\begin{equation}
	\frac{\partial \mathcal{L}_{tot}}{\partial \theta_o}
	= \frac{\partial \mathcal{L}_{mod}^o}{\partial \overline{O}} \cdot \frac{\partial \overline{O}}{\partial N_o} \cdot \frac{\partial N_o}{\partial \theta_o}
	+ \frac{\partial \mathcal{L}_{mod}^\rho}{\partial \overline{O}} \cdot \frac{\partial \overline{O}}{\partial N_o} \cdot \frac{\partial N_o}{\partial \theta_o}.
	\label{equ:b4}
\end{equation}
The first term originates from $\mathcal{L}_{mod}^o$. Since $\mathcal{L}_{mod}^o$ is defined directly as the sum of squared residuals of the differential equation for $\overline{O}$, its partial derivative with respect to $\overline{O}$, $\partial \mathcal{L}_{mod}^o / \partial \overline{O}$, is proportional to the residual itself (multiplied by a factor of 2). The second term originates from $\mathcal{L}_{mod}^\rho$, but the dependence of $\mathcal{L}_{mod}^\rho$ on $\overline{O}$ enters only through two terms in Eq.~(\ref{equ:1}): $[L,\rho_s\overline{O}^\dagger]$ and $[L^\dagger,\overline{O}\rho_s]$. Compared with the direct residual in $\mathcal{L}_{mod}^o$, the gradient signal from these terms propagates through more intermediate steps and lacks the quadratic amplification effect. Consequently, although $\partial \mathcal{L}_{mod}^\rho / \partial \theta_o  $ is mathematically non-zero, its typical numerical magnitude is far smaller than that of $\partial \mathcal{L}_{mod}^o / \partial \theta_o $. This means that the parameter update of the $\overline{O}$-branch is primarily driven by its own model loss $\mathcal{L}_{mod}^o$, while the cross-gradient from the $\rho$-branch serves only as a weak correction term.

For the $\rho$-branch parameters $\theta_\rho$, the chain-rule expansion is
\begin{equation}
	\frac{\partial \mathcal{L}_{tot}}{\partial \theta_\rho}
	= \; \frac{\partial \mathcal{L}_{mod}^\rho}{\partial \rho} \cdot \frac{\partial \rho}{\partial N_\rho} \cdot \frac{\partial N_\rho}{\partial \theta_\rho}
	+ \frac{\partial \mathcal{L}_{tar}}{\partial \rho} \cdot \frac{\partial \rho}{\partial N_\rho} \cdot \frac{\partial N_\rho}{\partial \theta_\rho} 
	+ \frac{\partial \mathcal{L}_{pos}}{\partial \rho} \cdot \frac{\partial \rho}{\partial N_\rho} \cdot \frac{\partial N_\rho}{\partial \theta_\rho}.
	\label{equ:b5}
\end{equation}
Here $\mathcal{L}_{mod}^\rho$ is the dominant driving term because it directly penalizes the residual of the entire density-matrix evolution equation, $\mathcal{L}_{tar}$ acts only at the terminal time, and $\mathcal{L}_{{pos}}$ merely enforces positive semi-definiteness. Both are auxiliary constraints whose gradient magnitudes are typically far smaller than that of $\mathcal{L}_{mod}^\rho$. Since $\mathcal{L}_{mod}^o$ does not depend on $\rho$ at all, there is no cross-gradient term from the $\overline{O}$ loss to $\theta_\rho$.

For the $c$-branch parameters $\theta_c$, both model losses depend on $N_c(t)$ through the Hamiltonian $H_s$
\begin{equation}
	\frac{\partial \mathcal{L}_{tot}}{\partial \theta_c}
	= \frac{\partial \mathcal{L}_{mod}^o}{\partial H_s} \cdot \frac{\partial H_s}{\partial N_c} \cdot \frac{\partial N_c}{\partial \theta_c}
	+ \frac{\partial \mathcal{L}_{mod}^\rho}{\partial H_s} \cdot \frac{\partial H_s}{\partial N_c} \cdot \frac{\partial N_c}{\partial \theta_c}.
	\label{equ:b6}
\end{equation}
The gradient of the $c$-branch naturally receives supervision from both dynamical equations simultaneously, enabling it to learn a driving field that satisfies both Eq.~(\ref{equ:1}) and Eq.~(\ref{equ:2}). The relative magnitudes of the two contributions depend on the sensitivity of each equation to the control field, but they are generally of comparable magnitude; therefore, the $c$-branch must balance the two objectives.

For the shared layer parameters $\theta_s$, the gradient originates from the sum of the gradients of all loss components after they have been backpropagated to the shared feature node $h_s$, and it is then propagated forward into the interior layers of the shared network. Let $h_s$ denote the output feature of the last shared layer (which also serves as the input feature for the three branches). Then
\begin{equation}
	\frac{\partial\mathcal{L}_{tot}}{\partial\theta_s}=\left(\frac{\partial\mathcal{L}_{mod}^o}{\partial h_s}+\frac{\partial\mathcal{L}_{mod}^\rho}{\partial h_s}+\frac{\partial\mathcal{L}_{tar}}{\partial h_s}+\frac{\partial\mathcal{L}_{pos}}{\partial h_s}\right)\cdot\frac{\partial h_s}{\partial\theta_s}.
	\label{equ:b7}
\end{equation}
The term in parentheses is precisely the sum of the gradients that the three branches send back to $h_s$. Since $h_s$ is the output of the shared layers, $\partial h_s /\partial \theta_s$ contains the derivatives of the weights in each interior layer of the shared network (transmitted layer by layer via the chain rule). Therefore, the shared layer parameters are updated through the aggregation of the gradients from all branch losses. Each branch's loss contributes to updating the shared layer parameters, which enables the shared layers to learn temporal features common to three tasks—simulating $\overline{O}$, $\rho$, and $c$. Moreover, because the gradients from different branches are directly summed at the $h_s$ node, even if the gradient direction of one branch is not perfectly aligned with those of the others, the shared layers are able to find a balanced update direction among the multiple objectives.

This gradient-flow structure embodies the core advantage of FPINN: although cross-gradients exist between branch layers due to physical coupling, the dominant update direction of each branch is still determined by its corresponding model loss, thereby effectively avoiding severe conflicts between different optimization objectives. The control branch acts as a ``bridge" between the two dynamical equations, naturally integrating supervisory information from both sides, which enables it to learn a driving field that simultaneously satisfies both dynamical constraints. The shared layers, through gradient accumulation, extract common temporal features from all branches. Compared with placing all outputs in a single network for training (where gradients are fully shared and objective conflicts readily arise), this design of partial isolation and selective sharing preserves the task specificity of each branch while fully exploiting the synergistic information embedded in the coupling relationships, thereby achieving a better balance between training stability and final accuracy.

\section*{References}

	\bibliographystyle{unsrt}
	\bibliography{References_library.bib}
	\clearpage

\end{document}